\documentstyle[12pt]{article}

\parskip=\medskipamount

\begin{document}

\vspace{8mm}

\begin{center}

{\Large \bf A note on the Faddeev-Popov determinant and Chern-Simons
perturbation theory} \\

\vspace{12mm}

{\large David H. Adams\footnote{Supported by FORBAIRT scientific research
program SC/94/218.}}

\vspace{4mm}

School of Mathematics, Trinity College, Dublin 2, Ireland. \\

\vspace{1ex}

email: dadams@maths.tcd.ie

\vspace{1ex}

June 1996

\end{center}

\begin{abstract}

A refined expression for the Faddeev-Popov determinant is derived for gauge 
theories quantised around a reducible classical solution. We apply this 
result to Chern-Simons perturbation theory on compact spacetime 3-manifolds
with quantisation around an arbitrary flat gauge field isolated up to
gauge transformations, pointing out that previous results on the finiteness
and formal metric-independence of perturbative expansions of the partition
function continue to hold.

\end{abstract}

\newpage

\noindent {\it Introduction}

The Faddeev-Popov gauge-fixing procedure \cite{FP}, used to rewrite the 
functional integrals arising in non-abelian gauge theories in a form to which
perturbative techniques can be applied, can be formulated in
the general setting where the theory is quantised around a general classical
solution \cite{AR}. 
However, when the classical solution is a reducible gauge field (e.g. the
zero-instanton in Yang-Mills theory on $S^4$ or an arbitrary flat gauge field 
in Chern-Simons theory on $S^3$ or lens spaces) the Faddeev-Popov determinant 
obtained by the standard derivation in this setting is degenerate, and the
perturbative techniques fail because the ghost propagator is ill-defined.
To avoid these and related problems the considerations have been restricted 
to irreducible gauge fields on many previous occasions in the literature
(e.g. \cite{BV(PLB)}, \cite{AxSi1}).
In this note we show that these problems can be avoided by a more careful 
derivation of the Faddeev-Popov determinant, taking account of certain
gauge-fixing ambiguities which arise when the classical solution is reducible.
The result is the following: Instead of the usual Faddeev-Popov determinant
\begin{eqnarray}
\det(\nabla_{\!\!A^c}^*\nabla_{\!\!A})\,,
\label{1}
\end{eqnarray}
where $A^c$ denotes the classical solution around which the theory is 
quantised, we obtain 
\begin{eqnarray}
V(H_{A^c})^{-1}\det\Bigl(\nabla_{\!\!A^c}^*\nabla_{\!\!A}
\biggl|_{\ker(\nabla_{\!\!A^c})^{\perp}}\,\Bigr)
\label{2}
\end{eqnarray}
Here $\nabla_{\!\!A}:\Gamma_{\!0}
\to\Gamma_{\!1}$ is the covariant derivative map determined by a gauge
field $A\,$; the gauge fields are identified with the connection 1-forms
on a principal fibre bundle $P$ over the compactified spacetime manifold
$M$ and $\Gamma_{\!q}$ denotes the space of q-forms on $M$ with values in
the bundle $P\times_{G}{\bf g}\,$, where the compact, semisimple gauge
group $G$ acts on its Lie algebra ${\bf g}$ by the adjoint representation.
The vectorspaces $\Gamma_{\!q}$ have inner products determined by a riemannian
metric on $M$ and invariant inner product in ${\bf g}\,$; these determine 
the adjoint map $\nabla_{\!\!A^c}^*$ in (\ref{1})-(\ref{2}) and the orthogonal
complement $\ker(\nabla_{\!\!A^c})^{\perp}$ 
of the nullspace $\ker(\nabla_{\!\!A^c})$ of $\nabla_{\!\!A^c}$.
Note that the determinant in (\ref{2}) makes sense at the formal level since
$\mbox{Im}(\nabla_{\!\!A^c}^*\nabla_{\!\!A})\subseteq\mbox{Im}
(\nabla_{\!\!A^c}^*)=\ker(\nabla_{\!\!A^c})^{\perp}$.
In (\ref{2}) $V(H_{A^c})$ denotes the (finite) volume of the 
isotropy group $H_{A^c}$ of
$A^c\,$, i.e. the subgroup of the group ${\cal G}$ of gauge transformations
which leave $A^c$ unchanged.

We go on to apply the result (\ref{2}) to the Chern-Simons perturbation theory
on compact spacetime 3-manifolds developed by S. Axelrod and I. Singer in
\cite{AxSi1}, pointing out that their results on the finiteness and formal
metric-independence of the perturbative expansions of the partition function
derived for $A^c$ an irreducible flat gauge field continue to hold for 
reducible $A^c$ when (\ref{2}) is used. As in \cite{AxSi1} we still require
$A^c$ to be isolated modulo gauge transformations though. The case of reducible
$A^c$ isolated up to gauge transformations is an important special case
since it applies for all the flat gauge fields on a number of basic 
3-manifolds, e.g. $S^3$ and the lens spaces, when $G=SU(2)$.

\vspace{1ex}

\noindent {\it The Faddeev-Popov determinant}

Recall that the Faddeev-Popov procedure for rewriting the functional 
integrals of the form
\begin{eqnarray}
\int{\cal D\/}A\,f(A)\,e^{-\frac{1}{\alpha^2}S(A)}
\label{x}
\end{eqnarray}
(where $S(A)$ is the action functional of the theory, $f(A)$ is a 
gauge-invariant functional and $\alpha$ is the coupling parameter\footnote{In
Chern-Simons gauge theory we replace $e^{-\frac{1}{\alpha^2}S(A)}$ in
(\ref{x}) by $e^{ikS(A)}$ where $S(A)$ is the Chern-Simons action 
functional.}) involves inserting $1=P_{A^c}(A)\,\Big/\,P_{A^c}(A)$ in the 
integrand, where
\begin{eqnarray}
P_{A^c}(A)=\int_{{\cal G}_0}{\cal D\/}\phi\,\delta(\nabla_{\!\!A^c}^*(\phi
{\cdot}A-A^c))
\label{x+1}
\end{eqnarray}
is the Faddeev-Popov functional associated with the gauge-fixing condition
\begin{eqnarray}
\nabla_{\!\!A^c}^*(A-A^c)=0
\label{x+2}
\end{eqnarray}
Following \cite{AR}, to avoid problems with the Gribov ambiguity,
we have taken the domain of the formal
integration in (\ref{x+1}) to be the subgroup ${\cal G}_0$ of topologically
trivial gauge transformations.
Using the ${\cal G}_0$-invariance of $S(A)\,$, $f(A)$ and 
$P_{A^c}(A)$ the resulting expression for (\ref{x}) is
\begin{eqnarray}
V({\cal G}_0)\,\int{\cal D\/}A\,f(A)\,e^{-\frac{1}{\alpha^2}S(A)}\,
P_{A^c}(A)^{-1}\,\delta(\nabla_{\!\!A^c}^*(A-A^c))
\label{x+3}
\end{eqnarray}
The standard evaluation of $P_{A^c}(A)^{-1}$ in (\ref{x+3})
leads to the Faddeev-Popov determinant (\ref{1}). We will show that a 
more careful evaluation of $P_{A^c}(A)^{-1}\,$ leads to the new
expression (\ref{2}).
The gauge-fixing condition (\ref{x+2}) has 
ambiguities coming from $H_{A^c}\,$, i.e.
\begin{eqnarray}
\nabla_{\!\!A^c}^*(A-A^c)=0
\ \ \Rightarrow\ \ \nabla_{\!\!A^c}^*(\phi{\cdot}A-A^c)=0
\qquad\ \ \ \forall\,\phi{\in}H_{A^c}\,.
\label{x+4}
\end{eqnarray}
To take this into account in the evaluation of $P_{A^c}(A)^{-1}$ we introduce
the map
\begin{eqnarray}
Q:\mbox{Lie}(H_{A^c})^{\perp}{\times}H_{A^c}\to{\cal G}_0\qquad,\qquad
Q(v\,,\phi):=\exp(v)\phi
\label{x+6}
\end{eqnarray}
where $\mbox{Lie}(H_{A^c})^{\perp}\subseteq\mbox{Lie}({\cal G}_0)=\Gamma_0$.
The differential (i.e. `Jacobi matrix') of $Q$ at $(0\,,\phi)\,$, 
\begin{eqnarray}
{\cal D\/}_{(0,\phi)}Q\,:\,\mbox{Lie}(H_{A^c})^{\perp}{\oplus}T_{\phi}H_{A^c}\,
\to\,T_{\phi}{\cal G}_0
\label{x+7}
\end{eqnarray}
is an isometry, so formally 
\begin{eqnarray}
|\det({\cal D\/}_{(0,\phi)}Q)|=1\ \ 
\ \ \ \ \ \ \ \ \ \ \forall\,\phi{\in}H_{A^c}\,.
\label{x+8}
\end{eqnarray}
(To see that (\ref{x+7}) is an isometry consider for fixed $\phi{\in}H_{A^c}$
the composition of maps
\begin{eqnarray}
\mbox{Lie}({\cal G}_0)=\mbox{Lie}(H_{A^c})^{\perp}\oplus\mbox{Lie}(H_{A^c})
\stackrel{\cong}{\longrightarrow}\mbox{Lie}(H_{A^c})^{\perp}{\oplus}
T_{\phi}H_{A^c}\stackrel{{\cal D\/}_{(0,\phi)}Q}
{\longrightarrow}T_{\phi}{\cal G}_0
\stackrel{\cong}{\longrightarrow}\mbox{Lie}({\cal G}_0) \nonumber 
\end{eqnarray}
where the first map is the isometry given by $(w,a)\mapsto
(w,\frac{d}{dt}\Bigl|_
{t=0}e^{ta}\phi)$ and the last map is the inverse of the isometry
$\mbox{Lie}({\cal G}_0)\stackrel{\cong}{\to}T_{\phi}{\cal G}_0$ given by
$v\mapsto\frac{d}{dt}\Bigl|_{t=0}e^{tv}\phi\,$. 
It is easy to see that this composition of maps is the identity on
$\mbox{Lie}({\cal G}_0)$.
It follows that ${\cal D\/}_{(0,\phi)}Q$ must be an isometry since all 
the other maps are isometries.) 
We now use the change of variables formula to calculate
\begin{eqnarray}
P_{A^c}(A)&=&\int_{{\cal G}_0}{\cal D\/}\phi\,
\delta(\nabla_{\!\!A^c}^*(\phi{\cdot}A-A^c)) \nonumber \\
&=&\int_{H_{A^c}{\times}\mbox{Lie}(H_{A^c})^{\perp}}
{\cal D\/}\phi{\cal D\/}v\,|\det({\cal D\/}_{(v,\phi)}Q)|
\,\delta\Bigl(\,\nabla_{\!\!A^c}^*(e^v\phi{\cdot}A-A^c)\Bigr) \nonumber \\
&=&\int_{H_{A^c}}{\cal D\/}\phi\,\Bigl|\det\biggl(\,
\nabla_{\!\!A^c}^*\nabla_{\!\!\phi{\cdot}A}
\biggl|_{\mbox{Lie}(H_{A^c})^{\perp}}
\,\biggr)\,\Bigl|^{-1} \nonumber \\
&=&\int_{H_{A^c}}{\cal D\/}\phi\,\Bigl|\det\biggl(\,
\nabla_{\!\!A^c}^*\nabla_{\!\!A}
\biggl|_{\mbox{Lie}(H_{A^c})^{\perp}}
\,\biggr)\,\Bigl|^{-1} \nonumber \\
&=&V(H_{A^c})\,\Bigl|\det\biggl(\,\nabla_{\!\!A^c}^*
\nabla_{\!\!A}\biggl|_{\mbox{Lie}(H_{A^c})^{\perp}}\,\biggr)\Bigl|^{-1}
\label{x+10}
\end{eqnarray}
where we have used (\ref{x+8}) in the second line and 
$\nabla_{\!\!A^c}^*\nabla_{\!\!\phi{\cdot}A}=
(\phi\cdot)\nabla_{\!\!A^c}^*\nabla_{\!\!A}(\phi\cdot)^{-1}$ 
for $\phi\in{}H_{A^c}$
in the third line.
Since $\nabla_{\!\!A^c}^*\nabla_{\!\!A^c}$ is a positive operator 
we can discard the numerical signs in
(\ref{x+10}) in the relevant case where $A$ is close to $A^c$ in ${\cal A}$
and arrive at the new expression (\ref{2}) for $P_{A^c}(A)^{-1}$ as promised.

The gauge-fixed functional integral (\ref{x+3}) can now be written in a form
which can be perturbativey expanded. We expand the action functional around
the classical solution $A^c$ as a polynomial in $B\in\Gamma_{\!1}\,$:
\begin{eqnarray}
S(A^c+B)=S(A^c)+<B\,,\,D_{\!A^c}B>+S_{A^c}^I(B)
\label{x+13}
\end{eqnarray}
where $D_{\!A^c}$ is a uniquely determined selfadjoint operator on $\Gamma_1$. 
Substituting the expression (\ref{2}) for $P_{A^c}(A)^{-1}$ in (\ref{x+3}) and
writing the determinant as a formal integral over independent anticommuting
variables $C\,,\,\bar{C}\in\ker(\nabla_{\!\!A^c})^{\perp}\,$
leads to the following
expression for the gauge-fixed functional integral:
\begin{eqnarray}
& &V({\cal G}_0)V(H_{A^c})^{-1}\det(\stackrel{\sim}{\Box}_{A^c})^
{-1/2}e^{-\frac{1}{\alpha^2}S(A^c)} \nonumber \\
&{\times}&\int_{\mbox{Im}(\nabla_{\!\!A^c})^{\perp}\oplus
\ker(\nabla_{\!\!A^c})^{\perp}
\oplus\ker(\nabla_{\!\!A^c})^{\perp}}{\cal D\/}({\alpha}
B){\cal D\/}\bar{C}{\cal D\/}C\,f(A^c+{\alpha}B)\,\exp\Big\{ \nonumber \\
& &-<B\,,D_{\!A^c}B>-<\bar{C}\,,\Box_{A^c}C>-\frac{1}{\alpha^2}S_{A^c}^I
({\alpha}B)-\alpha<\bar{C}\,,\nabla_{\!\!A^c}^*{\lbrack}B,C\rbrack>\Big\}
\nonumber \\
& &\label{x+14}
\end{eqnarray}
where $\Box_{A^c}=\nabla_{\!\!A^c}^*\nabla_{A^c}$ and
$\stackrel{\sim}{\Box}_{A^c}$ denotes its restriction to
the orthogonal complement of its nullspace. In the
Yang-Mills- and Chern-Simons gauge theories $\det(\stackrel{\sim}{\Box}_{A^c})$
can be given well-defined meaning via zeta-regularisation.
The ghost propagator $\Box_{A^c}^{-1}$
is well-defined on $\ker(\nabla_{\!\!A^c})^{\perp}$. The gauge field propagator
$D_{\!A^c}^{-1}$ on $\mbox{Im}(\nabla_{\!\!A^c})^{\perp}$ is well-defined
provided that $A^c$ is isolated up to gauge transformations in the space
${\cal C}$ of critical points for $S$.
This follows from $\ker(D_{\!A^c})=T_{A^c}{\cal C}$ (which can be shown by
a general argument when the moduli space of ${\cal C}$ is smooth at the 
point represented by $A^c$) since in this case 
$T_{A^c}{\cal C}=T_{A^c}({\cal G}{\cdot}A^c)=\mbox{Im}(\nabla_{\!\!A^c})$. 
Thus when $A^c$ is isolated up to gauge transformations the standard
perturbative techniques may be applied to (\ref{x+14}), leading to a 
perturbative expansion of the form
\begin{eqnarray}
V({\cal G}_0)Z_{sc}(\alpha;A^c)\sum_{V=0}^{\infty}\alpha^VI_V(A^c)
\label{x+15}
\end{eqnarray}
where $I_0(A^c)=f(A^c)$ and
\begin{eqnarray}
Z_{sc}(\alpha;A^c)=V(H_{A^c})^{-1}\,\det\Bigl(\frac{1}{\pi\alpha^2}
\tilde{D}_{\!A^c}\Bigr)^{-1/2}\,\det(\stackrel{\sim}{\Box}_{A^c})^{1/2}\,
e^{-\frac{1}{\alpha^2}S(A^c)}
\label{x+16}
\end{eqnarray}
The $V$'th term in (\ref{x+15}) is the contribution from all the Feynman
diagrams of order $V$. The weak coupling limit of (\ref{x+15}),
$V({\cal G}_0)Z_{sc}(\alpha;A^c)f(A^c)\,$, coincides with the contribution
from $A^c$ to the semiclassical approximation for (\ref{x}) obtained from
the invariant integration method of A.~Schwarz \cite[App. II (9)]{Sch(Inst)}.
This is reassuring since Schwarz's method does not use gauge fixing, unlike
ours. This also indicates that (\ref{2}) will allow
the relationship between the Faddeev-Popov
determinant and the natural metric on the orbit space of gauge fields, pointed
out by O.~Babelon and C.-M.~Viallet \cite{BV(PLB)} when
the considerations are restricted to irreducible gauge fields,
to be extended to the reducible case, although we will not pursue this here.

\vspace{1ex}

\noindent
{\it Application to Chern-Simons perturbation theory on compact 3-manifolds}

Following \cite{AxSi1} we consider perturbative expansion of the 
Chern-Simons partition function
\begin{eqnarray}
Z(M,k)=\frac{1}{V({\cal G}_0)}\int{\cal D\/}A\,e^{ikS(A)}
\label{y}
\end{eqnarray}
where 
\begin{eqnarray}
S(A)=\frac{1}{4\pi}\int_{M}\mbox{Tr}(A{\wedge}dA+\frac{2}{3}A
{\wedge}A{\wedge}A)\,.
\label{y+1}
\end{eqnarray}
Here the spacetime $M$ is a closed oriented 3-manifold, and for simplicity
$P$ is assumed trivial so the gauge 
fields $A$ are ${\bf g}$-valued 1-forms on $M$. 
Let $A^c$ be an arbitrary flat gauge field on $M$ which
is isolated up to gauge transformations, with flat covariant derivative
$d=d^{A^c}=\oplus_{q=0}^3d_q^{A^c}$ on the space $\Omega=\Omega(M,{\bf g})=
\oplus_{q=0}^3\Omega^q(M,{\bf g})$ of ${\bf g}$-valued differential forms 
on $M$ (so $\nabla_{\!\!A^c}=d_0^{A^c}$), 
and let $H(A^c)=\oplus_{q=0}^3H^q(A^c)$
denote its cohomology. The requirement that $A^c$ be isolated up to gauge 
transformations is equivalent to $H^1(A^c)=0$.

We apply the gauge-fixing procedure of the preceding section to (\ref{y}).
In this case we obtain the perturbative expansion
\begin{eqnarray}
Z(M,k,A^c)=Z_{sc}(M,k,A^c)\sum_{V=0,2,4,\dots}\Bigl(\frac{1}{\sqrt{k}}
\Bigr)^VI_{V}(M,A^c)
\label{y+2}
\end{eqnarray}
where $I_0(M,k,A^c)=1$ and
\begin{eqnarray}
Z_{sc}(M,k,A^c)=e^{-\frac{i\pi}{4}\eta({\ast}d_1)}\,\Bigl(\frac{4\pi
\lambda_{\bf g}}{k}\Bigr)^{{\dim}H^0(A^c)/2}\,
V(H_{A^c})^{-1}\,\tau(M,A^c)^{1/2}
\label{y+3}
\end{eqnarray}
In obtaining (\ref{y+3}) we have used the results of \cite{AdSe(PLB)}.
Here $<a,b>_{\bf g}=-\lambda_{\bf g}\mbox{Tr}(ab)$ and $\tau(M,A^c)$ is the 
Ray-Singer torsion of $A^c$. The modulus of (\ref{y+3}) is metric-independent
\cite[\S5]{Sch(degen)}. The metric-dependent phase factor is discussed
in \cite[\S2]{W(Jones)}. 
It has been verified \cite{semicl} for wide classes of 3-manifolds that
the expression for the semiclassical approximation for $Z(M,k)$ obtained
from (\ref{y+3}) coincides with the weak coupling (i.e. large $k$) limit
of the expressions for $Z(M,k)$ obtained from Witten's non-perturbative
prescription \cite{W(Jones)}.

In \cite{AxSi1} the perturbative expansion (\ref{y+2}) was considered for
irreducible $A^c$. 
The results of the preceding section allow these considerations to be
extended to reducible $A^c$ isolated up to gauge transformations
(after suitable changes of variables in (\ref{x+14}))
and the expressions for the
coefficients $I_V(M,A^c)$ derived in \cite{AxSi1} continue to hold:
$I_V(M,A^c)=0$ for $V$ odd, and for $V$ even \cite[(3.54)--(3.55)]{AxSi1}
\begin{eqnarray}
I_V(M,A^c)&=&c_V\prod_{i=1}^V\bigg\lbrack\int_{M_{x_i}}f_{a^ib^ic^i}
\frac{\partial}{{\partial}j_{(i)}^{a^i}}
\frac{\partial}{{\partial}j_{(i)}^{b^i}}
\frac{\partial}{{\partial}j_{(i)}^{c^i}}\bigg\rbrack\,L_{tot}
(x_1,\dots,x_V)^{\frac{3}{2}V} \nonumber \\
&=&c_V\int_{M^V}\mbox{TR}\Bigl(L_{tot}(x_1,\dots,x_V)^{\frac{3}{2}V}\Bigr)
\label{y+4}
\end{eqnarray}
where $c_V=(2{\pi}i)^{\frac{1}{2}V}((3!)^V(2!)^{\frac{3}{2}V}V!
(\frac{3}{2}V)!)^{-1}$. 
Briefly, the notations are as follows 
(see \cite[\S2--3]{AxSi1} for the details)\footnote{Here and throughout 
this section all repeated indices are to be summed over.}. 
$\{j^a\}$ is an o.n.b. for ${\bf g}\,$, ${\lbrack}j^a,j^b\rbrack=f_{abc}j^c\,$,
$\frac{\partial}{{\partial}j_{(i)}^a}$ is interior
multiplication by $j_{(i)}^a\,$,
$L_{ab}(x,y)\in\Omega^2(M_x{\times}M_y)$ (singular at $x=y$) is given by
\begin{eqnarray}
(\hat{L}\psi)^a(x)=\int_{M_y}L_{ab}(x,y)\wedge\psi^b(y)\qquad\qquad
\psi=\psi^bj^b\in\Omega(M,{\bf g})
\label{y+5}
\end{eqnarray}
where $\hat{L}:\Omega(M,{\bf g})\to\Omega(M,{\bf g})$ is given by
\begin{eqnarray}
d\hat{L}=\pi_d\qquad\;\;\qquad\;\;\hat{L}d=\pi_{d^*}
\label{y+6}
\end{eqnarray}
$\pi_d$ and $\pi_{d^*}$ are the projections onto the images of 
$d$ and $d^*\,$, and finally 
\begin{eqnarray}
L_{tot}(x_1,\dots,x_V)&=&\sum_{i,k=1}^VL_{ab}(x_i,x_k){\wedge}j_{(i)}^a
{\wedge}j_{(k)}^b \label{y+7} \\
&{\in}&\Gamma(M_{x_1}\times\cdots{\times}M_{x_V};
\Lambda(\oplus_{i=1}^V(T^*M_{x_i}\oplus{\bf g}_i)))\,. \nonumber
\end{eqnarray}

We conclude this note by pointing out that the finiteness-- and formal 
metric--independence properties of the perturbative expansion (\ref{y+2})
derived in \cite{AxSi1} for irreducible $A^c$ continue to hold in the 
present context. 
With the point--splitting regularisation of \cite{AxSi1}
the expression (\ref{y+4}) for $I_V(M,A^c)$ is finite 
\cite[theorem 4.2]{AxSi1}. The argument for this goes through for arbitrary
flat $A^c$ \cite[\S6 Remark II(i)]{AxSi1}.
The extension of the metric--independence result of \cite{AxSi1} to the 
present case is less obvious. It was shown in \cite[\S5]{AxSi1} that the
expression (\ref{y+4}) for $I_V(M,A^c)$ is formally metric--independent
for irreducible $A^c$. (A subsequent rigorous treatment 
\cite[\S5]{AxSi1} \cite{AxSi2}, taking account of the singularities in
$L_{tot}$ in the diagonals $x_i=x_k\,$,
reveals an anomalous metric--dependent phase). It was known
to the authors of \cite{AxSi1} that the formal metric--independence of
$I_V(M,A^c)$ continues to hold for reducible $A^c$ 
\cite[\S6 Remark II(i)]{AxSi1}, but since an argument has not previously
been provided in the literature we will give one here. A rigorous treatment 
of the problem in the very general case where $A^c$ is only required to
belong to a smooth component of the modulispace has recently been announced
and outlined by S.~Axelrod \cite{Ax(hep-th)}. However, the powerful new
algebraic techniques outlined there are not necessary to show the formal 
metric--independence in the present case. As we will see, this can be shown 
by much simpler means.

To show the formal metric-independence of (\ref{y+4}) in the present case
we need generalisations of the properties of the propagator derived in
\cite[\S3]{AxSi1}. The property (PL1) generalises to
\begin{eqnarray}
d_{M_x{\times}M_y}L_{ab}(x,y)=(d_{M_x}+d_{M_y})L_{ab}(x,y)=
\Bigl(\delta_{ab}\delta(x,y)-\pi_{ab}(x,y)\Bigr)
\label{y+8}
\end{eqnarray}
where $\pi_{ab}(x,y)$ is obtained from the projection $\pi$ onto the harmonic
forms in the same way that $L_{ab}(x,y)$was obtained from $\hat{L}$
(see (\ref{y+5})). Explicitly\footnote{We are following the convention
of \cite[(3.53)]{AxSi1}.},
\begin{eqnarray}
\pi_{ab}(x,y)=\sum_ih_i^a(x)h_i^b(y)V(M)^{-1}(vol(y)-vol(x))
\label{y+9}
\end{eqnarray}
where $vol(x)$ and $vol(y)$ are the volume forms on $M_x$ and $M_y$ 
respectively, considered as elements in $\Omega^3(M_x{\times}M_y)\,$, and
$\{h_i=h_i^aj^a\}$ is a metric-independent basis for $\ker(d_0)$
chosen so that $<h_i(x),h_k(x)>_{\bf g}=\delta_{ik}$ ${\forall}x{\in}M$.
We decompose
\begin{eqnarray}
L_{ab}(x,y)&=&L_{ab}^{(0,2)}(x,y)+L_{ab}^{(1,1)}(x,y)+L_{ab}^{(2,0)}(x,y)
\label{y+10} \\
\pi_{ab}(x,y)&=&\pi_{ab}^{(0,3)}(x,y)+\pi_{ab}^{(3,0)}(x,y) \label{y+11}
\end{eqnarray}
where $Q^{(p,q)}(x,y)\in\Omega(M_x{\times}M_y)$ denotes a form of degree
p on $M_x$ and degree q on $M_y$. Using (\ref{y+8}) we find that the
generalisation of the key property (PL4) of \cite[\S3]{AxSi1} is
\begin{eqnarray}
\delta_{{\delta}g}L_{ab}^{(1,1)}(x,y)=d_{M_x{\times}M_y}B_{ab}(x,y)
\label{y+12}
\end{eqnarray}
for some $B(x,y)\in\Omega^1(M_x{\times}M_y,{\bf g}\otimes{\bf g})$ of the
form
\begin{eqnarray}
B_{ab}(x,y)=B_{ab}^{(0,1)}(x,y)-B_{ba}^{(1,0)}(x,y)
\label{y+13}
\end{eqnarray}
together with
\begin{eqnarray}
d_{M_x}\Bigl(\delta_{{\delta}g}L_{ab}^{(0,2)}(x,y)\Bigr)=0\qquad\quad,
\qquad\quad\,d_{M_y}\Bigl(\delta_{{\delta}g}L_{ab}^{(2,0)}(x,y)\Bigr)=0\,.
\label{y+14}
\end{eqnarray}
Here ${\delta}g$ is a variation of the chosen metric $g$ on $M$.

Now, repeating the calculation \cite[(5.83)]{AxSi1} gives in the present
case
\begin{eqnarray}
\delta_{{\delta}g}I_V(M,A^c)=-\frac{3}{2}V(\frac{3}{2}V-1)c_V(I_V^{(1)}-
I_V^{(2)})+3Vc_VI_V^{(3)}
\label{y+16}
\end{eqnarray}
where
\begin{eqnarray}
I_V^{(1)}&=&\int_{M^V}\mbox{TR}\Bigl(B_{tot}^{(0,1)}\delta_{tot}^{\bf g}
(L_{tot})^{\frac{3}{2}V-2}\Bigr)
\label{y+17} \\
I_V^{(2)}&=&\int_{M^V}\mbox{TR}\Bigl(B_{tot}^{(0,1)}\pi_{tot}
(L_{tot})^{\frac{2}{3}V-2}\Bigr)
\label{y+18} \\
I_V^{(3)}&=&\int_{M^V}\mbox{TR}\Bigl((\delta_{{\delta}g}L_{tot}^{(0,2)})
(L_{tot})^{\frac{3}{2}V-1}\Bigr)
\label{y+19}
\end{eqnarray}
(The derivation uses Stoke's theorem, and is therefore formal since
$L_{tot}(x_1,\dots,x_V)$ is not smooth on $M^V$. At all other points here
and below we are rigorous). The integral $I_V^{(1)}$ is the one appearing
in the calculation of \cite[\S5]{AxSi1}, and vanishes by the argument
given there. The integrals $I_V^{(2)}$ and $I_V^{(3)}$ are new features
of the present, more general situation where $A^c$ is reducible.
The key to showing that they vanish is to note that
\begin{eqnarray}
\int_Mf_{abc}h^a\phi^b\wedge\psi^c=0\qquad\qquad\forall\,h\in\ker(d_0)\;,\;
\phi\in\ker(d^*)\;,\;\psi\in\mbox{Im}(d^*)
\label{y+20}
\end{eqnarray}
and
\begin{eqnarray}
L_{ab}(x,y)\in\mbox{Im}(d_{M_x}^*)\qquad\qquad,\qquad\qquad\,
L_{ab}(x,y)\in\mbox{Im}(d_{M_y}^*)
\label{y+21}
\end{eqnarray}
The formula (\ref{y+20}) follows from $\int_M\phi^a\wedge\psi^a=0$
$\forall\phi\in\ker(d^*)\,,\,\psi\in\mbox{Im}(d^*)$ together with
${\lbrack}h,\psi\rbrack\in\mbox{Im}(d^*)$ ${\forall}h\in\ker(d_0)\,,\,
\psi\in\mbox{Im}(d^*)\,$, while (\ref{y+21}) follows from (\ref{y+5})
and (\ref{y+6}).
To see that $I_V^{(3)}$ vanishes note that it can be expanded as a sum
of terms where each term involves an integral of the form
\begin{eqnarray}
\int_{M_y}f_{ace}(\delta_{{\delta}g}L_{ab}^{(0,2)}(y,x_i))L_{cd}(y,x_j))
L_{ef}(y,x_k)
\label{7.32}
\end{eqnarray}
(There are also terms where $L_{cd}(y,x_j)L_{ef}(y,x_k)$ is replaced by
$L_{ce}(y,y)$ in (\ref{7.32}) but these vanish since the integrand 
contains no 3-forms in $y$ in this case). 
Because of (\ref{y+14}) and (\ref{y+21}) it follows from (\ref{y+20})
that (\ref{7.32}) vanishes.

The argument for the vanishing of $I_V^{(2)}$ is slightly more involved.
Note that $\pi_{tot}=\pi_{tot}(x_1,\dots,x_V)$ is given by a sum of terms
as in (\ref{y+7}), leading to an expression for $I_V^{(2)}$ as a sum of 
corresponding terms, each consisting of an integral over $M^V$. A number
of these terms vanish for one of the following reasons: \hfil\break
(i) $\pi_{ab}(x,x)=0$. (This follows from (\ref{y+9})). \hfil\break
(ii) The integrand in the integral over $M^V$ (a differential form on $M^V$)
is not of degree 3 in $x_i$ for all $i=1,\dots,V$. (Then the integral over 
$M_{x_i}$ vanishes). \hfil\break
(iii) The term contains an integral of the form
\begin{eqnarray}
\int_{M_y}f_{abd}h^a(y)L_{bc}(y,x_i)L_{de}(y,x_j)
\label{7.33}
\end{eqnarray}
which vanishes by (\ref{y+20})--(\ref{y+21}).
The only terms which do not
vanish due to (i), (ii) or (iii) are those of the form
\begin{eqnarray}
\lefteqn{\int_{M_z{\times}M_{x_i}{\times}M_{x_j}{\times}M_{x_k}{\times}M_y}
\Big\{
f_{acd}f_{bfp}h^a(y)B_{bc}^{(0,1)}(z,y)L_{de}^{(2,0)}(y,x_i)} \nonumber \\
& &\qquad\qquad\times\,L_{fg}(z,x_j)L_{pq}(z,x_k)\Psi_{egq}(x_i,x_j,x_k)\Big\}
\label{7.34}
\end{eqnarray}
or
\begin{eqnarray}
\lefteqn{\int_{M_z{\times}M_{x_i}{\times}M_{x_j}{\times}M_y}
\Big\{f_{ade}f_{bcg}
h^a(y)h^b(z)vol(z)B_{cd}^{(0,1)}(z,y)} \nonumber \\
& &\qquad\qquad\times\,L_{ef}^{(2,0)}(y,x_i)L_{gh}(z,x_j)\Phi_{fh}(x_i,x_j)
\Big\}
\label{7.34.5}
\end{eqnarray}
To show that these vanish it suffices to show that
\begin{eqnarray}
\int_{M_y}f_{acd}h^a(y)B_{bc}^{(0,1)}(z,y)L_{de}^{(2,0)}(y,x_i)
\in\ker((d_{M_z})_0)
\label{7.35}
\end{eqnarray}
Then (\ref{7.34})--(\ref{7.34.5}) vanish due to (\ref{y+20})--(\ref{y+21})
(note for (\ref{7.34.5}) that $h(z)vol(z)\in\ker(d_{M_z}^*)$ ).
To show (\ref{7.35}) we begin by noting that 
\begin{eqnarray}
\int_{M_y}f_{acd}h^a(y)L_{bc}^{(1,1)}(z,y)L_{de}^{(2,0)}(y,x_i)=0
\label{7.36}
\end{eqnarray}
for the same reason that (\ref{7.33}) vanished in (iii) above. Taking the 
metric-variation of this gives 
\begin{eqnarray}
0&=&\int_{M_y}f_{acd}h^a(y)(\delta_{{\delta}g}L_{bc}^{(1,1)}(z,y))L_{de}^
{(2,0)}(y,x_i)+\int_{M_y}f_{acd}h^a(y)L_{bc}^{(1,1)}(z,y)\delta_{{\delta}g}
L_{de}^{(2,0)}(y,x_i) \nonumber \\
&=&d_{M_z}\int_{M_y}f_{acd}h^a(y)B_{bc}^{(0,1)}(z,y)L_{de}^{(2,0)}(y,x_i)
+\int_{M_y}f_{acd}h^a(y)L_{bc}^{(1,1)}(z,y)\delta_{{\delta}g}L_{de}^{(2,0)}
(y,x_i) \nonumber \\
& &\label{7.37}
\end{eqnarray}
where we have used (\ref{y+12})--(\ref{y+13}). The first term in (\ref{7.37})
belongs to $\mbox{Im}(d_{M_z})$ while the second term belongs to
$\mbox{Im}(d_{M_z}^*)$ because of (\ref{y+21}). Since $\mbox{Im}(d^*)=
\ker(d)^{\perp}\subseteq\mbox{Im}(d)^{\perp}$ it follows that both terms in
(\ref{7.37}) vanish individually; the vanishing of the first term implies
(\ref{7.35}). This completes the argument for the formal metric--independence
of $I_V(M,A^c)$.

\vspace{2ex}

\noindent {\it Conclusion}

We carried out a more careful version of the Faddeev-Popov gauge-fixing 
procedure for gauge theories quantised around a reducible classical solution
$A^c\,$, finding a new refined expression (\ref{2}) for the Faddeev-Popov 
determinant. Unlike the usual expression when $A^c$ is reducible, the 
ghost propagator associated with this expression is well-defined. When $A^c$
is isolated up to gauge transformations in the space of classical solutions
the gauge field propagator is also well-defined, and in this case the standard
perturbative techniques can be applied to the gauge-fixed functional integrals.
We applied this to the partition function of Chern-Simons gauge theory on
a general compact 3-manifold, showing that the previous results of 
S.~Axelrod and I.~Singer on the finiteness-- and formal metric--independence
of the perturbation series continue to hold for reducible $A^c$.
This opens up the possibility of carrying out explicit perturbative
expansions of the Chern-Simons partition function for $S^3$ and lens spaces.
Our expression for the lowest order term is consistent with
the previously calculated nonperturbative
expressions \cite{semicl} for the weak coupling (large $k$) limit of
the partition function. 

\vspace{1ex} 

{\it Acknowledgements.} 
I am grateful to Siddhartha Sen for valuable discussions and encouragement
during the course of this work.

\end{document}